# Novelty and Social Search in the World Wide Web

Bernardo A. Huberman and Lada A. Adamic
Xerox Palo Alto Research Center
3333 Coyote Hill Road
Palo Alto, CA 94304

## Abstract

The World Wide Web is fast becoming a source of information for a large part of the world's population. Because of its sheer size and complexity users often resort to recommendations from others to decide which sites to visit. We present a dynamical theory of recommendations which predicts site visits by users of the World Wide Web. We show that it leads to a universal power law for the number of users that visit given sites over periods of time, with an exponent related to the rate at which users discover new sites on their own. An extensive empirical study of user behavior in the Web that we conducted confirms the existence of this law of influence while yielding bounds on the rate of novelty encountered by users.

The World Wide Web (Web) has become in a very short period one of the most useful sources of information for a large part of the world's population. Its exponential growth, from a few sites in 1994 to millions today, has transformed it into an ecology of knowledge in which highly diverse information is linked in extremely complex and arbitrary fashion(1). In addition, the rapid growth of electronic commerce is bringing about a further increase in the amount of information available on the Web.

A problem with this bounty is that it is often hard for users to choose which sites to visit, if only because it is difficult to judge a priori their information value. In addition, since providers of electronic commerce sometimes lack recognizable reputations and can offer similar services, it is seldom possible to make optimal decisions as to which sites to access and which ones to avoid. As with many other situations were choice is costly, people resort to a cooperative mechanism which relies on the collective search performed by a whole community to find desirable and useful sites. This is an extremely successful strategy, which has been shown to lead to faster problem solving(2), increased genetic variability among certain species(3) and speedier diffusion of innovations in many social settings(4).

In the case of the Web, social search can also solve the problem posed by its explosive growth, for it is impossible for a single individual or the best search engines to keep track of all the newly created sites and the informational value that they contribute. Large groups of people surfing on their own can sample a much larger information space than single individuals, and any exchange of relevant findings can increase the awareness of possibly interesting sites. Even though recommendations, both personal and institutional(5-7), can be unreliable and highly idiosyncratic, they decrease the cost of searching for optimal sources of information, while leading to the discovery of new sites and improved ways of surfing the Web.

This social search mechanism is particularly useful when looking for useful information, and should be contrasted with surfing, in which users follow hyperlinks as long as they find value in them(8). Users typically use recommendations to choose a starting site, while their activity within the site is determined by their surfing behavior. Thus one expects that the mechanism of social search will be manifested in the statistics of site visits of the Web.

In what follows we study the dynamics of recommendations and their effect on site visits by users of the World Wide Web. We show that this form of social search leads to a universal power law for the number of users that visit given sites over periods of time. The exponent of the power law is related to the rate at which users discover new sites on their own, as opposed to following recommendations. An extensive empirical study of user behavior in the Web that we conducted confirms the existence of this law of influence while yielding bounds on the rate of novelty encountered by users.

Consider the familiar situation of Internet surfers having to choose among a large number of Web sites. Given the unstructured character of the Web, users will either surf on their own or follow recommendations from others to visit useful sites. Notice that the nature of these recommendations can range from messages suggesting visits to particular sites, to brand recognition and links to other sites that appear in pages that individuals are visiting. If people in turn find value in those sites, they might recommend them to others. Recommendations will thus lead individuals to preferentially choose those sites that have been already selected by other individuals. This probabilistic process determines the number of users that visit given sites. In addition, there is an important novelty factor, since users can visit sites that have not been previously recommended by using search engines that do not explicitly rank their links so as to bias users to particular sites. This is an important discovery mechanism for newly created sites that have either few visitors or small number of links to them.

In order to understand the dynamics of site visits as a function of both site discovery and recommendations, it is useful to formulate this problem in terms of a probabilistic urn model(9). If one identifies sites with colors and visitors to those sites as balls of that color, a user following a recommendation can be thought of as blindly extracting a ball from an urn, noting its color, and replacing it in the urn along with another ball of the same color. Since in this urn model the probability that a user will select a site is proportional to the number of previous visitors to the site, the addition of the user to the site's visitors makes subsequent selection of the site more probable. In addition, the process of discovery of new sites can be thought of as the occasional random insertion of a new color into the urn.

This urn mechanism can be formulated as a branching process(10) by stating that at a given time step the probability that a site visited by a user has not been recommended is $\nu$, so that if $s$ is the total number of sites seen at time $t$, $P(s \to s+1; t \to t+1) = \nu$. On the other hand, with probability $1-\nu$ a user will select a site previously viewed and recommended by someone else, with the probability of being selected proportional to the number of users, $n_s$, that have visited it previously, i.e., $P(n_s \to n_s+1; t \to t+1) = (1-\nu)n_s t^{-1}$, with $t = \sum_{s=1}^{S} n_s$.

The evolution of the system is described by the master equations for the probability that at time t, a total of s novel sites have been discovered, and that the s site has been visited by $n_s$ users. In terms of time steps it is given by

$$P(s+1, t+1) = \nu P(s,t) + (1-\nu)P(s+1,t) \tag{1}$$

$$P(n_s+1, t+1) = (1-\nu)n_s t^{-1} P(n_s,t) + \nu P(n_s+1,t) + \nu P(s-1,t)\delta(n_s,0) \tag{2}$$

with the last term of Eq. (2) expressing the constraint that the *s* site is discovered at time *t*. Notice that since at any given time step either a site is either discovered or visited following a recommendation, the value of the novelty factor $\nu$ is bounded away from one.

The solution to these equations gives the probability that $n$ users visit a site, and depends on the time of discovery of that site. Since Web sites appear at different times, the probability that over a long period of time a site has been visited by $n$ new users, regardless of the time of its appearance, is a mixture of those distributions.

This mixture is obtained by summing the individual site probabilities, which are geometrically distributed, weighted by an exponential factor reflecting the growth of the Web. As originally shown by Yule in the context of speciation (11, 12), the resulting distribution is given by

$$P(n) = \frac{(\alpha-1)\Gamma(\alpha)\Gamma(n)}{\Gamma(n+\alpha)} \tag{3}$$

where $\Gamma(x) = \int_0^\infty u^{x-1} e^{-u} du$, and the coefficient $\alpha$ is given in terms of the novelty coefficient $\nu$ by

$$\alpha(\nu) = \frac{2-\nu}{1-\nu} \tag{4}$$

which is shown in Fig. 1 for a range of values of $\nu$. Since for large values of $n$, $\Gamma(n)/\Gamma(n+\alpha) \approx n^{-\alpha}$ we can write Eq. (3) as,

$$P(n) = \Gamma(\alpha) n^{-\alpha} \tag{5}$$

which implies that, up to a constant given by the Gamma function, the probability of finding $n$ users having visited a given site scales inversely in proportion to their number. Thus, the law of influence predicts that when plotted on a log-log plot, the data shows a straight line whose slope is determined by the novelty coefficient through Eq. (4).

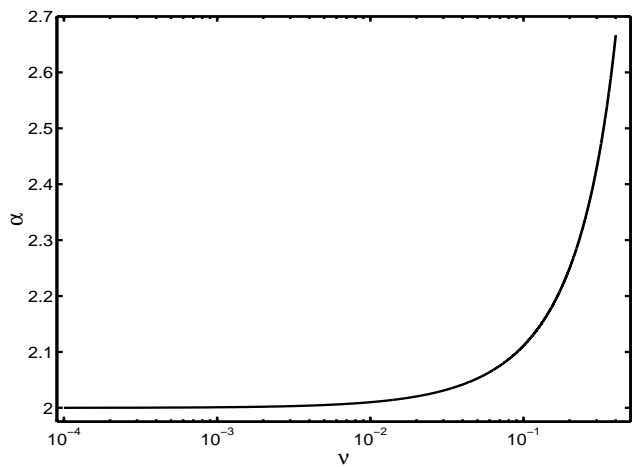

Fig.1 The relationship between $\nu$, the novelty coefficient and $\alpha$, the power law exponent.

In order to test the theory we conducted an extensive analysis of several Web client traces containing data about site visits for thousands of sites. The first set of data consisted of America On-Line (AOL) traces obtained separately for the dates of December 3rd and 5th, 1997. On December 3rd 47,399 users visited 93,838 sites, while on December 5th 23,316 users visited 87,430 sites. We assume that on a given day users already familiar with a site will revisit it with some probability and that new users will visit it in proportion to the number of users that have already visited the site. Thus, an analysis of trace data over a window of time should yield the relative fraction of users frequenting each site. In the case of AOL, we extracted the number of unique visitors a site received by removing multiple visits by the same user.

Figure 2 shows the resulting distribution on a log-log scale for the data of Dec. 3rd, 1997. As can be seen, it exhibits the power law predicted by Eq. (5) with an exponent $\alpha = 2.03 \pm 0.05$. Using Eq. (4) this corresponds to a value of the novelty factor $\nu < 0.1$. which implies that out of ten users visiting new sites at least nine were following recommendations. Regression analysis yields goodness of fit parameters $R^2 = 0.998$, $P < 10^{-6}$, confirming the ability of the recommendation model to account for user statistics. Similarly, a log-log plot of the data for Dec. 5th, yields a straight line, with a similar exponent, which implies the same bound on $\nu$.

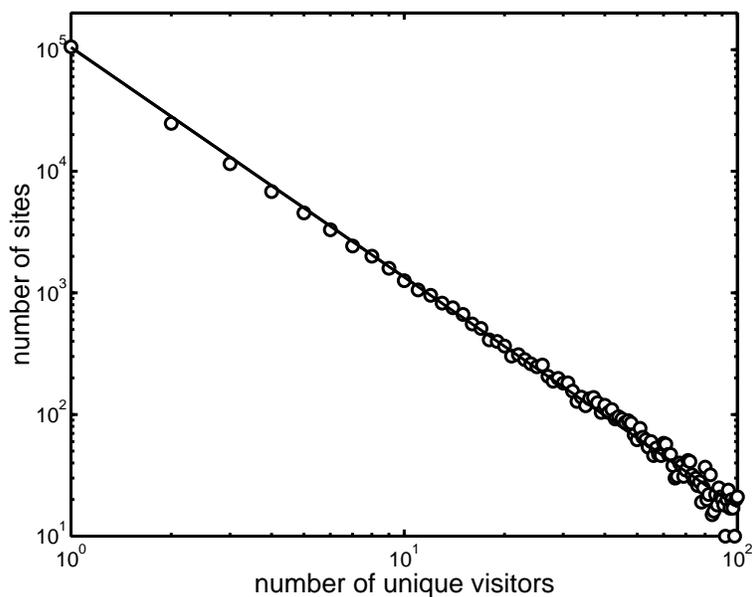
Fig.2 Distribution of unique AOL visitors per site on Dec. 3, 1997.

In order to test the universality of the theory we also analyzed an older set of trace data collected by Cunha et al. at Boston University (13). It contains traces of 558 students who accessed 2,225 sites during the months of January and February of 1995. Once again, as shown in Figure 3, one obtains power law behavior, with a value of the exponent $\alpha = 2.08 \pm 0.08$, corresponding to a novelty rate $< 0.2$.

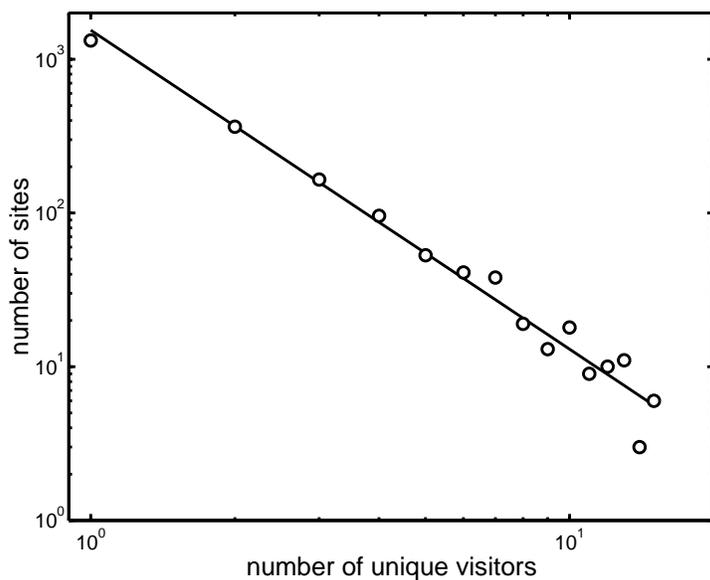
Fig. 3 Distribution of unique visitors per site from BU traces Jan.-Feb. 1995.

The difference in value between the exponent corresponding to the Boston University and AOL data can be attributed to two different factors. The first one has to do with the fact that the novelty rates among the sets of users in the two cases are different, whereas the second relates to the length of time sampled by the data. Consider two sites that have the same number of unique visitors but are visited at different frequencies-one might be a news site users access daily, the other an online travel agent accessed only once a month. If one studies the usage logs for the two sites during a single day, one will appear to have more visitors than the other. Thus, data covering a shorter period of time would have the effect of broadening the distribution given by Eq. (5) and therefore yield a smaller exponent value. Since the AOL data covers only a single day of Web usage, it has a smaller value of $\alpha$ than the Boston University data, which was collected over two months and hence is less sensitive to the frequency of access to sites.

Another interesting result is obtained if the site visit data is plotted as a function of the site rank, instead of number of unique visitors. As shown in Figure 4, a Zipf-like law (14) is observed, with an exponent whose value is 0.9. This confirms ealier findings by Glassman (15) and Aida et al. (16). We point out however, that Zipf-like regularities do not provide useful information about the particular process that generates them. This is because the process of ranking random variables stemming from any broad distribution always produces a narrow and monotically decreasing power law of the type originally discussed by Zipf (17).

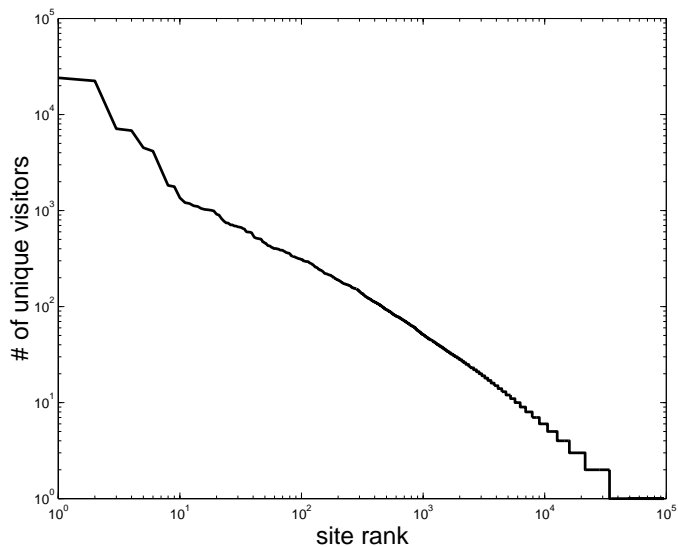

Fig.4 Ranked order distribution of the number of unique visitors per site.

These results show that a theory of social search in the context of the World Wide Web can account for the strong statistical regularities observed in site visits. A universal power law, whose exponent relates to the rate at which individuals discover new Web sites, describes these regularities. We conclude by pointing out that the sheer size of the Web and its highly interactive nature makes it provides a fertile ground for the discovery and testing of laws of social behavior that are harder to verify in other settings.

**Acknowledgements**

We thank Jim Pitkow for providing the data for our analysis and his comments on the paper, and Rajan Lukose for many useful discussions. This work was partly supported by NSF Grant IRI-961511.

**References**

1. Special issue on the Internet, Scientific American **276** (1997).
2. S. H. Clearwater, B. A. Huberman and T. Hogg, Cooperative Solution of Constraint Satisfaction Problems, Science **254**, 1181-1183 (1991).
3. M. J.Wade and S.G. Pruett-Jones, Female Copying Increases the Variance in Male Mating Success, Proc. Natl. Acad. Sci. USA, **87**, 5749-5753 (1990).
4. E. M .Rogers. Diffusion of Innovations. Free Press (1983).
5. http://www.alexa.com
6. W. C. Hill, L. Stead, M. Rosenstein and G. Furnas, Recommending and Evaluating Choices in a Virtual Community of Use, in Proceedings of CHI'95, 194-201 ACM Press (Denver, CO), 1995
7. N. Glance, D. Arregui and M. Dardenne, Knowledge Pump: Supporting the Flow and Use of Knowledge, in Information Technology for Knowledge Management. Eds. U. Borghoff and R. Pareschi, Springer-Verlag, 1998
8. B. A. Huberman, P. Pirolli, J. Pitkow and R. M. Lukose, Strong Regularities in World Wide Web Surfing, Science **280**, 95-97 (1998).
9. N. L. Johnson and S. Kotz, Urn Models and their Application (John Wiley), (1977).
10. K. B. Athreya and P. E. Ney, Branching Processes, (Springer-Verlag, 1972).
11. G. U. Yule, A Mathematical Theory of Evolution Based on the Conclusions of Dr. J. C. Willis, Philosophical Transactions of the Royal Society of London, Series B, **213**, 21-87 (1924).
12. N. L. Johnson, S. Kotz, A.W. Kemp, Univariate Discrete Distributions (John Wiley & Sons) (1992).
13. C. R. Cunha, A. Bestavros and M. E. Crovella, Characteristics of WWW Client-based Traces, Dept. of Computer Science, Boston University Report (1995).
14. G. K. Zipf, Human behavior and the principle of least effort (Addison-Wesley, Cambridge, MA, 1949
15. S. Glassman, Computer Networks and ISDN Systems **27**, 165-173 (1994).
16. M. Aida and N. Takahashi, Evaluation of the Number of Destination Hosts for Data Networking, Proceedings of the Sixth International Conference on Computer Communications and Networks, IEEE 342-349 (1997).
17. R.Gunther, L. Levitin, B. Schapiro and P. Wagner, Zipf's Law and the Effect of Ranking on Probability Distributions, International Journal of Theoretical Physics, 35, 395-417 (1996).

.